\documentclass[a4paper,11pt]{article}
\usepackage{pos}

\usepackage{cleveref}
\usepackage{caption}
\usepackage{subcaption}
\usepackage{wrapfig}

\title{Estimation of Aperture of the Tunka-Rex Radio Array for Cosmic-Ray Air-Shower Measurements}
 \ShortTitle{Estimation of the Tunka-Rex aperture}

\author*[a]{Vladimir Lenok}
\affiliation[a]{Karlsruhe Institute of Technology, Institute for Astroparticle Physics, D-76021 Karlsruhe, Germany}

\forColl{Tunka-Rex} 

\emailAdd{contact@tunkarex.info}

\abstract{The recent progress in the radio detection technique for air
  showers paves the path to future cosmic-ray radio detectors. Digital
  radio arrays allow for a measurement of the air-shower energy and
  depth of its maximum with a resolution comparable to those of the
  leading optical detection methods. One of the remaining challenges
  regarding cosmic-ray radio instrumentation is an accurate estimation
  of their efficiency and aperture. We present a probabilistic model
  to address this challenge. We use the model to estimate the
  efficiency and aperture of the Tunka-Rex radio array. The basis of
  the model is a parametrization of the radio footprint and a
  probabilistic treatment of the detection process on both the antenna
  and array levels. In this way, we can estimate the detection
  efficiency for air showers as function of their arrival direction,
  energy, and impact point on the ground. In addition, the transparent
  internal relationships between the different stages of the
  air-shower detection process in our probabilistic approach enable to
  estimate the uncertainty of the efficiency and, consequently, of the
  aperture of radio arrays. The details of the model will be presented
  in the contribution.}

\FullConference{37$^{\rm{th}}$ International Cosmic Ray Conference (ICRC 2021)\\
		July 12th -- 23rd, 2021\\
		Online -- Berlin, Germany}

\begin{document}
\maketitle

\section{Introduction}

Aperture estimation for a cosmic-ray radio antenna array is an open
problem in modern ground-based observational astroparticle
physics. This work presents a probabilistic model of the detection
efficiency combined with a semi-analytic way of the aperture
estimation, which both form the aperture model. The model was
developed for the Tunka-Rex instrument; however, it can be applied to
other cosmic-ray radio instruments.

\subsection{Tunka-Rex Instrument}

The Tunka Radio Extension (Tunka-Rex) is a cosmic-ray digital radio
antenna array located at the site of the TAIGA facility in the Tunka
valley in Siberia~\cite{2015NIMPA.802...89B, 2019JPhCS1263a2006K}. The
antenna array evolved and consisted of 63 short aperiodic loaded loop
antennas (SALLA~\cite{2012JInst...7P0011A}) distributed over about
1~km\textsuperscript{2} by the operation's end in 2019. The antennas'
operating frequency band ranges from 30 to 80~MHz.

\subsection{Definition of Aperture}

Aperture as a physical quantity naturally appears in consideration of
the number, $N$, of cosmic rays observed by an instrument. This number
equals the cosmic-ray flux $J$ multiplied to exposure $\epsilon$ of
the instrument, namely, $N = \epsilon\,J$. The exposure is the crucial
factor for the reconstruction of the cosmic-ray flux's characteristics
from the observations. Its generic form can be expressed as the
integral of the efficiency $\xi$ over the area of fiducial angular
selection, $\Omega_f$, the instrumental fiducial area, $S_f$, and the
duration of the measurements, $T$
\begin{equation}
  \epsilon =
  \int_{T} \int_{\Omega_f} \int_{S_f}
  \xi ~ \cos{\theta} \, ds \, do \, dt.
\end{equation}
If the efficiency does not constantly change, the
time-independent part of the integral can be considered
separately. This part is referred to as ``aperture''
\begin{equation}
  A =
  \int_{\Omega_f} \int_{S_f}
  \xi ~ \cos{\theta} \, ds \, do.
  \label{eq-aperture-generic}
\end{equation}
This integral can be transformed
\begin{multline}
  \int_{\Omega_f} \int_{S_f}
  \xi ~ \cos{\theta} \, ds \, do =
  S_f \, \int_{\Omega_f} \int_{S_f}
  \frac{\xi}{S_f} ~ \cos{\theta} \, ds \, do = \\
  S_f \, \int_{\Omega_f}
  \left(\frac{1}{S_f} \int_{S_f} \xi \, ds\right) ~ \cos{\theta} \, do =
  S_f \, \int_{\Omega_f}
  \langle \xi \rangle_s ~ \cos{\theta} \, do =  S_f A_{\Omega}.
  \label{eq-aperture}
\end{multline}
The symbol $\langle \xi \rangle_s$ denotes the averaged efficiency
over the fiducial area. The last integral denoted as $A_{\Omega}$ is
referred to as ``angular aperture''\footnote{ In the same fashion, the
  aperture could be factorized into a solid angle of the fiducial
  angular selection and the averaged efficiency over this angle.  Even
  though this approach can have advantages in some studies, since it
  preserves the angular coverage of the instrument in case of
  selection of the full-efficiency regions, it is out of the scope of
  the present work.
}.
Thus, by collecting the introduced quantities, the exposure estimation
can be expressed in the factorized form
\begin{equation}
  \epsilon = S_f \, A_{\Omega} \, T.
\end{equation}

\section{Efficiency Model}

As it is clear from the definition above, efficiency is the critical
component in aperture estimation. The usual approach of the efficiency
estimation with extensive libraries of the Monte-Carlo simulations is
cumbersome in studying air-shower radio emission since the
corresponding simulations are computationally expensive. We use a
different approach for the aperture study --- we build a model for the
instrument efficiency and use it in the aperture estimation.

The idea behind the model is to exploit the probabilistic properties
of air-shower radio footprint characteristics and the properties of
the air-shower detection process. The model consists of three
components: the model of air-shower radio footprint
(section~\ref{sec-footprint}), the model of signal detection on the
level of individual antennas (section~\ref{sec-antenna-level}), and
the model of entire-array triggering (section~\ref{sec-trigger}).

\subsection{Spatial Distribution of Radio Signals}
\label{sec-footprint}
The Tunka-Rex asymmetric lateral distribution function in the
discussed model is the means of predicting the spatial distribution of
the electric field, $\mathscr{E}$, corresponding to a given set of the
air-shower parameters, i.e., incoming direction, energy, $E$, depth of
shower maximum, $X_{\text{max}}$. The reconstruction uses the same
function~\cite{2019ICRC...36..331L}.

The model describes the distribution in the geomagnetic coordinate
system, the system built on the basis of the shower propagation vector
$\textbf{V}$ and the local geomagnetic field $\textbf{B}$ vector.

The lateral distribution function shows only the most probable value
of the electric field. The distribution of the electric field strength
for a given spatial point is an inseparable component of the model.
We model this distribution with the normal distribution with a
standard deviation of 14.47\% of the field value centered on the most
probable value. The value of the standard derivation comes from a
direct comparison of the model predictions and the Monte-Carlo
simulated signals.

\subsection{Signal Detection by Individual Antennas}
\label{sec-antenna-level}

The process of signal detection by individual antennas has a
probabilistic nature because of the unavoidable presence of
noise. Even though this nature is especially pronounced for small
signals, it significantly influences the detection process for a wide
range of signals relevant for the air-shower radio array operation.

\begin{figure}[!t]
    \begin{subfigure}{0.5\textwidth}
    \centering
  \includegraphics{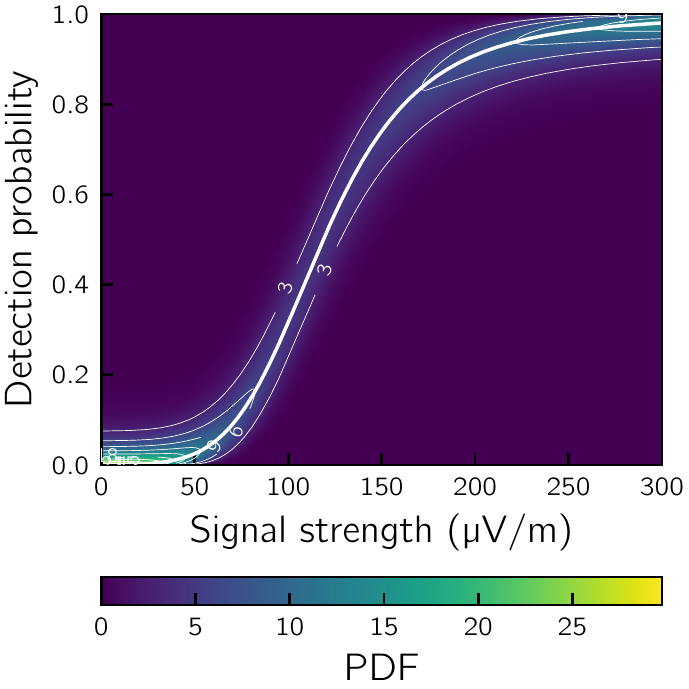}
\end{subfigure}
\begin{subfigure}{0.5\textwidth}
\centering
  \includegraphics{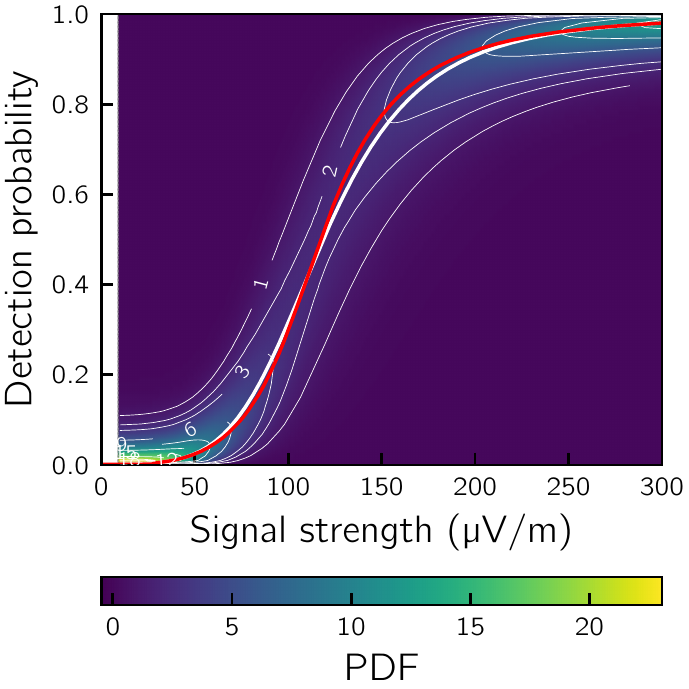}
\end{subfigure}
    \caption{The probability density function of the signal detection
      on the individual antenna level. \textit{Left}:~initial
      function. The white line shows the most probable
      value. \textit{Right}:~the function after the marginalization
      procedure taking into account the uncertainty of the signal
      description. The white and red lines indicate the most probable
      values for the function before and after the marginalization.}
    \label{fig-detectability-pdf}
\end{figure}

To describe this probabilistic behavior, we processed the Monte-Carlo
generated electric field with the Tunka-Rex signal-processing pipeline
\cite{Tunka-Rex:2015zsa}, which includes the antenna model, response
of the electronic components, and signal analysis procedure. The
processing was done multiple times with different samples of the
on-site measured noise samples. The fraction of times the standard
analysis pipeline of Tunka-Rex could detect a given signal becomes the
detection probability for this signal. The S-shaped curve parametrizes
the obtained dependence of the detection probability $p_0$ as a
function of the signal strength $S$
\begin{equation}
  p_0(S) = \frac{1}{2} +
  \frac{1}{2}\tanh \frac{S-S_{1/2}}{S'_0 + S''_0 S}.
  \label{eq-s-curve}
\end{equation}
Within the framework of the model, the detection probability is not a
number but a random variable characterized by the probability density
of the beta distribution form
\begin{equation}
    P = \frac{p^{\alpha-1}\left(1-p\right)^{\beta-1}}
        {\mathrm{B}(\alpha,\beta)}
  \label{eq-prob-det}
\end{equation}
with the parameters depending on the most probable value
$\alpha = n p_0+ 1$, $\beta = n - n p_0 + 1$,
and the number of repetitions of the signal processing, $n$, that
equals 30 for the present work. The left plot in
Figure~\ref{fig-detectability-pdf} shows the obtained probability
density.

To account for the uncertainty of the signal description, we apply a
marginalization procedure. The function under the marginalization is a
combination of the probability density of signal
detection~(\ref{eq-prob-det}) and the probability density of observing
a signal with a certain strength modeled with the normal distribution,
$\mathcal{N}$, with the standard deviation equal to 14.47\% found from
the comparison of the model and the simulations,
$\sigma_S = 0.1447\,S$.
The combined density is subject to marginalization over the signal
strength
\begin{equation}
   P_S(S)= \int_0^{\infty}
   \frac{p^{\alpha(S')-1}\left(1-p\right)^{\beta(S')-1}}
   {\mathrm{B}(\alpha(S'),\beta(S'))} \,
   \mathcal{N}(S, \sigma^2_S(S'))\, dS'.
\end{equation}

The marginalization makes the probability density wider, allowing for
a broader range of the possible realizations of the detection
probability. The right plot in Figure~\ref{fig-detectability-pdf}
shows the obtained marginalized function.

\subsection{Fulfillment of Trigger Condition}
\label{sec-trigger}

This stage combines the probability densities obtained on the
individual antenna level into a final probability density used to
infer the detection probability value for a given shower. The entire
Tunka-Rex array triggering happens when at least a certain number of
antennas detect signals. Two methods can provide probability estimation
for such triggering scheme.

\pagebreak
\begin{wrapfigure}{r}{0.5\textwidth}
  \includegraphics{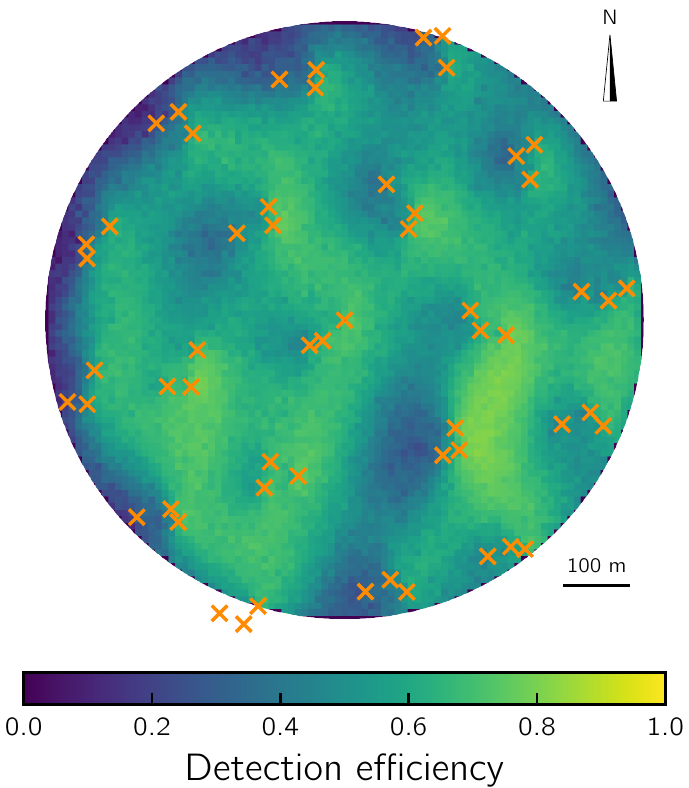}
  \caption{The detection efficiency as a function of the air-shower
    core location for a shower with the following parameters
    $\theta$~=~35$^{\circ}$, $\phi$~=~270$^{\circ}$,
    $\log_{10}(E/ \text{1\,eV})$~=~17.3, X$_{\text{max}}$~=~658 g/cm$^2$.
    The outer circle depicts the border of the instrument fiducial
    area. The sign in the upper right corner shows the direction towards
    the geographic north.}
  \vspace{-10mm}
  \label{fig-fieldmap}
\end{wrapfigure}
\paragraph{Probabilistic Calculations.}
One way to estimate the probability of observing a given number of
signals over the antenna field is to perform the computation of
probabilities of all possible situations not leading to the
triggering and to subtract them from unity
\begin{equation}
  P = 1 - \sum_{i=1}^{\binom{N}{0}} p^{(0)}_i -
  \sum_{i=1}^{\binom{N}{1}} p^{(1)}_i - \cdots -
  \sum_{i=1}^{\binom{N}{m-1}} p^{(m-1)}_i .
  \label{eq-prob-sum}
\end{equation}
Each term is a sum of all possible configurations leading to the
observation of zero signals, one signal, etc., till $m-1$, where $m$
is the least number of signals required for triggering. In their
explicit form, the terms look like the products of the corresponding
probability density functions, $p$ for the probability to detect a
signal and $\bar{p} = 1-p$ for the probability of the non-detection,
the index indicates the particular antenna
\begin{equation}
\begin{aligned}
    p^{(1)} = &p_1 \bar{p}_2 \cdots \bar{p}_{n-1} \bar{p}_n + \\
    &\bar{p}_1 p_2 \cdots \bar{p}_{n-1} \bar{p}_n + \cdots + \\
    &\bar{p}_1 \bar{p}_2 \cdots \bar{p}_{n-1} p_n.
\end{aligned}
\end{equation}
This equation gives an example for the probability where only a single
antenna has a signal. Each term in the sum is a possible combination
of the antennas with and without signals. The total number of the
terms equals the total number of the possible combinations described
by the corresponding binomial coefficient. In this particular case it
is $\binom{N}{1}$.

A numerical procedure estimates the probability functions in practical
calculations. The peak, the most probable value, of the obtained
probability density function is the detection probability for a given
air shower. Figure~\ref{fig-fieldmap} shows the detection probability
for multiple air-shower core positions.

This calculation method provides an accurate estimation of the
probability density function of the trigger fulfillment. However, its
computational complexity rises very fast with the increasing number of
antennas required for triggering. The second method, presented below,
does not feature such behavior.

\paragraph{Monte-Carlo experiments.}
Another way to estimate the probability density is to run Monte-Carlo
experiments. The idea behind this method is an extension of the
existing idea of estimating the trigger probability by running a
Bernoulli process for each detector with a given signal-detection
probability. In every run, the Bernoulli process on each detector
randomly assigns detection or non-detection flag to the detector. By
running this procedure multiple times and computing the fractions of
configurations leading to satisfying the trigger condition, one can
estimate the detection probability for an array. In the most simple
case, described here, the trigger condition is simply the number of
detectors with signals.

Our new method extends this idea from the probabilities to the
probability densities. We draw one random sample value from each
probability density to detect a signal by an individual antenna and
run the procedure described above to obtain a value of the trigger
probability. Then the procedure repeats for different sample values of
the probability densities. The obtained sample of the trigger
probabilities forms the probability density function. The most
probable value of the obtained function is the array trigger
probability for a given air shower.

\begin{figure}[!t]
    \begin{subfigure}{0.5\textwidth}
    \centering
  \includegraphics{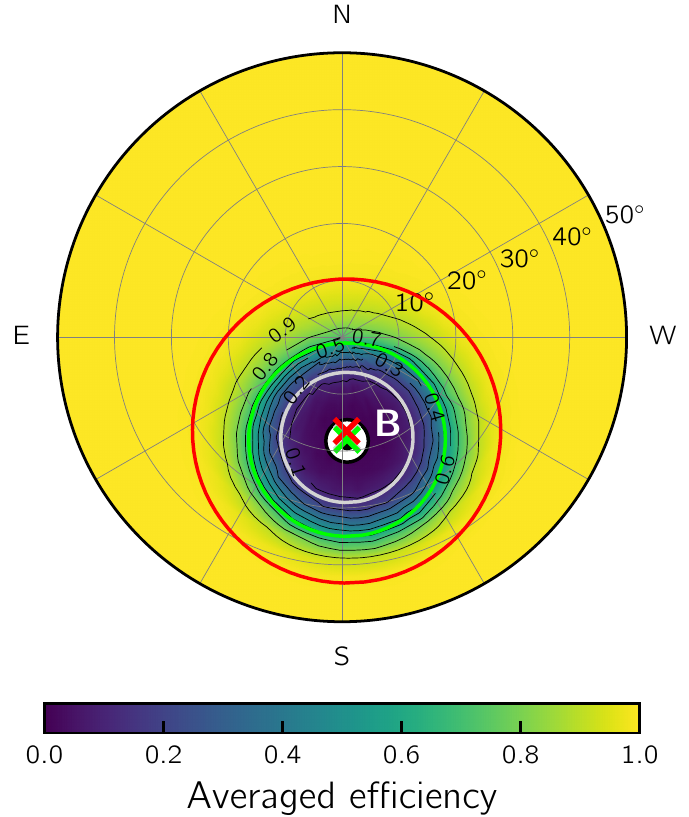}
\end{subfigure}
\begin{subfigure}{0.5\textwidth}
\centering
  \includegraphics{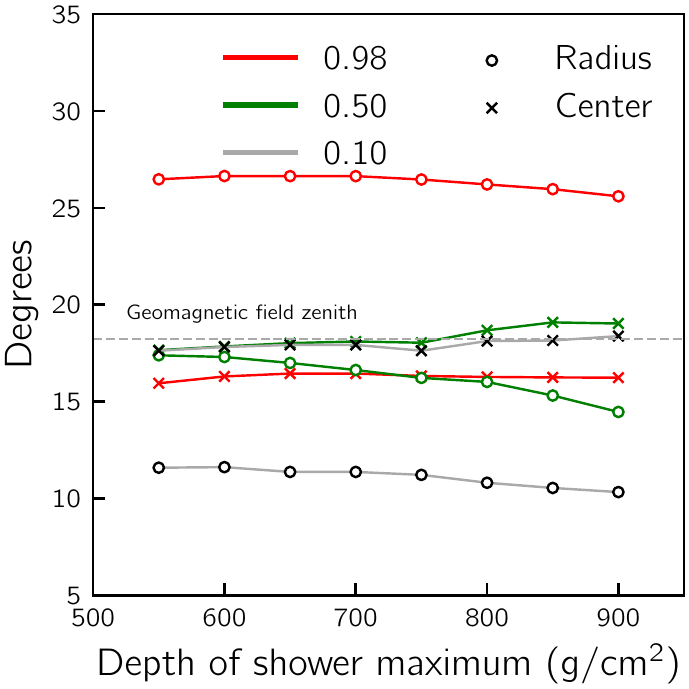}
\end{subfigure}
    \caption{The angular behavior of the averaged efficiency.
    \textit{Left}: the distribution of the averaged efficiency over
    the sky. The red, green, and gray circles correspond to the 0.98,
    0.5, and 0.1 maximal efficiency regions.
    \textit{Right}: the evolution of the radii and center positions of
    the circles corresponding to the 0.98, 0.50, and 0.1 maximal
    efficiency regions. The size of the 0.98 efficiency circle is
    almost independent of X$_{max}$.}
    \label{fig-circles}
\end{figure}

\section{Estimation of Aperture}

According to the definition of the angular aperture introduced in the
equation~(\ref{eq-aperture}), its value is the projected averaged
efficiency integrated over the sky. The averaged efficiency features
quite complex angular behavior, directly impacting the aperture
estimation for a radio array. Figure~\ref{fig-circles} shows the
averaged efficiency for Tunka-Rex resulting from the efficiency model
described above. As expected from the physics of air-shower radio
emission, the efficiency has a suppressed region around the direction
of the geomagnetic field.

\subsection{Selection of Full-Efficiency Region}

Even though the behavior of the detection efficiency over the sky is
complex, the selection of only the full-efficiency part can be made
relatively simple. The region of the suppressed efficiency close to
the geomagnetic field direction can be approximated with a circle on a
sphere.  The size of the circular region covering the suppressed
efficiency region almost does not depend on the
X$_{\text{max}}$. Figure~\ref{fig-circles} shows the circular region
parameters obtained for a given air-shower energy of 10$^{17.3}$~eV
and several positions of X$_{\text{max}}$ ranging from 550 to
900~g/cm$^2$. As one can see the sizes of the 98\% and 10\% efficiency
regions effectively remain the same, however, the size of the 50\%
efficiency region undergo significant change.  The behavior of the
later region is likely connected to the increase of the brightness of
the shower while closing the distance to the observer. The exact
explanation of the overall angular behavior of the efficiency will be
investigated.


\subsection{Evaluation of the Aperture Integral}

\begin{wrapfigure}{r}{0.5\textwidth}
\vspace{-10mm}
  \includegraphics{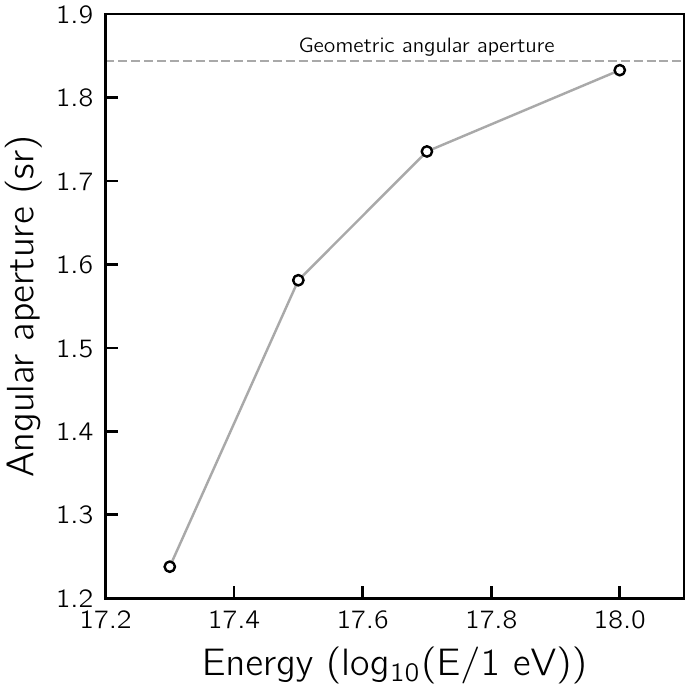}
  \caption{The angular aperture for Tunka-Rex array in the full
    configuration of 63 antennas and the default trigger condition, at
    least three antennas with signals. \\
    (Reference X$_{max}$~=~650~g/cm$^2$.)}
  \label{fig-apperture}
  \vspace{-10mm}
\end{wrapfigure}

It is possible to evaluate the aperture integral within the full
efficiency region semi-analytically. Namely, express the
two-dimensional integral in a one-dimensional analytical form, which,
in turn, can be computed numerically with very high precision.

The starting point for the integration is to put the averaged
efficiency, $\langle \xi \rangle_s$, equal to unity. For convenience,
we use the spherical coordinate system where the integral appears in
its usual form
\begin{multline}
  A_{\Omega} =
  \int_{\Omega_f}
  \langle \xi \rangle_s ~ \cos{\theta} \, do = \\
  \int_0^{2\pi} \int_0^{\theta_{max}} \cos\theta \sin\theta \, d\theta \, d\phi.
\end{multline}

The whole-sky area within the maximal zenith angle is referred to as
region I. The region within the circular area with lower efficiency is
referred to as region II, the area within the red circle in
Figure~\ref{fig-circles}. The integral is evaluated for all directions
within region I, and then the value of the integral for region II is
subtracted. The solutions for these two integrals are known, the
details of the derivation are in
Reference~\cite{2019JPhCS1181a2027L}. The solution for the aperture
has the form
\begin{multline}
  A_{\Omega} = (I) - (II) =
  \pi \left[ 1 - \cos^2\theta_{max} \right] - \\
  2 \int_{0}^{\theta_{max}}
  \arccos
  \left( \frac{\cos\rho - \cos\theta \cos\theta_0}{\sin\theta \sin\theta_0}
  \right)
  \cos\theta \sin\theta \,
  d\theta,
\end{multline}
where the letter $\theta_0$ denotes the zenith location of the region
II, the letter $\rho$ denotes the angular radius of region II. Factor
2 appears due to the symmetry of the circular region.

Figure~\ref{fig-apperture} shows the estimation of aperture for
discrete energies in the range from 10\textsuperscript{17.3} to
10\textsuperscript{18.0}~eV. The size of the suppressed-efficiency
region shrinks at the highest energies, making the instrumental
aperture almost equal to the geometrical aperture.

\section{Conclusions and Outlook}

The presented model of the aperture estimation combines two major
components: the probabilistic model of the detection efficiency and
the semi-analytical approach to the aperture integral estimation. The
model incorporates all known uncertainties and uses all available
information about the underlying probability distributions. The
model's accuracy limitations come only from the accuracy of the
suppressed region exclusion and the accuracy of the numerical
integration.

With this model it is possible to study the energy and
X$_{\text{max}}$ dependence of the efficiency suppression
region. Figure~\ref{fig-circles} shows only the first study of this
kind, indicating that the size and location of the
efficiency-suppressed region are almost constant over a wide range of
X$_{\text{max}}$.

The model has a generic nature and can be used in a wide range of
applications. It can be adopted to other air-shower radio arrays with
appropriate selection of the footprint description and an update of
the signal detection probability.

{\small
\section*{Acknowledgment}
The present work on the Tunka-Rex project is funded by Deutscher
Akademischer Austauschdienst~e.V. (personal grant, ref. number
91657437, program ID 57299294).  In preparation of this work we used
calculations performed on the computational resource ForHLR II funded
by the Ministry of Science, Research and the Arts
Baden-W\"{u}rttemberg and DFG (“Deutsche Forschungsgemeinschaft”). A
part of the data analysis was performed using the radio extension of
the Offline framework developed by the Pierre Auger
Collaboration~\cite{PierreAuger:2011btp}.

\bibliographystyle{ICRC}
\bibliography{references}
}



%
%
%

\clearpage
\section*{Full Authors List: \Coll\ Collaboration}
\scriptsize
\noindent
P.~Bezyazeekov$^{1}$,
N.~Budnev$^{1}$,
O.~Fedorov$^{1}$,
O.~Gress$^{1}$,
O.~Grishin$^{1}$,
A.~Haungs$^{2}$,
T.~Huege$^{2,3}$,
Y.~Kazarina$^{1}$,
M.~Kleifges$^{4}$,
E.~Korosteleva$^{5}$,
D.~Kostunin$^{6}$,
L.~Kuzmichev$^{5}$,
V.~Lenok$^{2}$,
N.~Lubsandorzhiev$^{5}$,
S.~Malakhov$^{1}$,
T.~Marshalkina$^{1}$,
R.~Monkhoev$^{1}$,
E.~Osipova$^{5}$,
A.~Pakhorukov$^{1}$,
L.~Pankov$^{1}$,
V.~Prosin$^{5}$,
F.~G.~Schr\"oder$^{2,7}$
D.~Shipilov$^{8}$ and
A.~Zagorodnikov$^{1}$
~\\
~\\
\noindent
$^{1}$Applied Physics Institute ISU, Irkutsk, 664020 Russia\\
$^{2}$Karlsruhe Institute of Technology, Institute for Astroparticle Physics, D-76021 Karlsruhe, Germany\\
$^{3}$Astrophysical Institute, Vrije Universiteit Brussel, Pleinlaan 2, 1050 Brussels, Belgium\\
$^{4}$Institut f\"ur Prozessdatenverarbeitung und Elektronik, Karlsruhe Institute of Technology (KIT), Karlsruhe, 76021 Germany\\
$^{5}$Skobeltsyn Institute of Nuclear Physics MSU, Moscow, 119991 Russia\\
$^{6}$DESY, Zeuthen, 15738 Germany\\
$^{7}$Bartol Research Institute, Department of Physics and Astronomy, University of Delaware, Newark, DE, 19716, USA\\
$^{8}$X5 Retail Group, Moscow, 119049 Russia

\end{document}